\documentclass[runningheads]{svmult}

\usepackage{makeidx}   
\usepackage{graphicx}  
\usepackage{subeqnar}  
\usepackage{multicol}  
\usepackage{physprbb}  
\makeindex             

\begin{document}
\title*{Spectral Properties of Short Gamma-Ray Bursts}
\toctitle{Spectral Properties of Short Gamma-Ray Bursts}
%
%
\titlerunning{Spectral Properties of Short GRBs}
%
\author{W.S. Paciesas\inst{1}
\and R.D. Preece\inst{1}
\and M.S. Briggs\inst{1}
\and R.S. Mallozzi\inst{1,2}}
\authorrunning{W.S. Paciesas et al.}
%
%
\institute{University of Alabama in Huntsville, AL 35899, USA
\and deceased}

\maketitle              

\begin{abstract}

The distribution of GRB durations is bimodal, but there is little additional
evidence to support the division of GRBs into short and long classes. Based on
simple hardness ratios, several studies have shown a tendency for longer GRBs
to have softer energy spectra.  Using a database of standard model fits to
BATSE GRBs, we compare the distributions of spectral parameters for short and
long bursts. Our preliminary results show that the average spectral break
energy differs discontinuously between short and long burst classes, but within
each class shows only a weak dependence on burst duration.

\end{abstract}

Various studies have shown that short and long GRBs are statistically different
classes~\cite{dezalay92,kouvel93,belli95,mukherjee98,horvath98}. Recently,
additional evidence has come from a study by Norris et al~\cite{nor3}, who
found no measurable energy-dependent pulse lag in the time histories of short
events. This is in contrast to long GRBs, which clearly show such a
lag~\cite{nor2}, even for short subpulses. Moreover, in bursts with measured
redshifts (which thus far are all long events), the energy-dependent pulse lag
appears to be anti-correlated with burst luminosity~\cite{nor2}. Thus, if the
mechanism producing the lag works for short bursts, they must be intrinsically
more luminous than long bursts, and therefore more distant. Alternatively, a
different mechanism may operate in short events. Either way, the evidence seems
to support separate classification of short and long GRBs.

In the currently
favored fireball model, the prompt burst emission is thought to be optically
thin synchrotron or synchrotron self-Compton emission from internal shocks, as
external shocks are unable to produce the observed temporal structure
\cite{fenimore96,saripiran97}. Detailed studies of the spectra of a number of
bright GRBs, including both long and short events, have shown good consistency
with the synchrotron shock model~\cite{tavani96a,tavani96b,cohen97}. However,
more comprehensive analyses have uncovered problems with this
interpretation~\cite{crider97,preece98b,preece01}. In particular, some GRB
spectra are harder at low energies than the synchrotron
limit~\cite{crider97,preece98b}. These conclusions are based mostly on
spectroscopy of long bursts, but the problem may be most acute for short bursts
because their spectra are on average harder. 

Recent work~\cite{preece98a,preece00} has characterized the range of spectral
behavior in bright, long bursts in some detail, but the spectral properties of
the class of short bursts have only been characterized using hardness ratios. 
Phebus data showed that short GRBs are harder than long ones~\cite{dezalay92}
(confirmed by many succeeding analyses of BATSE data), but detailed study of
the spectral differences between short and long bursts has not been done. In
particular, the consistency of short burst spectra with the synchrotron shock
model predictions has not been properly tested. It is clear that a better
characterization of the spectral differences between short and long bursts is
warranted. For the foreseeable future, the BATSE data base will provide the
best sample of bursts for this purpose. 

The BATSE {\it CONT} datatype is derived from the large area detectors and is
independent of the BATSE trigger. However, the {\it CONT} data have 16-channel
energy resolution and 2~s time resolution, so they are not optimal for the
analysis of the spectra of short events because the 2~s integration degrades
the signal-to-noise ratio. Nevertheless, a database of {\it CONT} fits was
conveniently available~\cite{mallozzi98}, so we used these data to perform a
preliminary study of spectral differences between short and long GRBs. The {\it
CONT} fit database contains spectral fits for $\sim$1200 BATSE GRBs. Fit
results for two spectra per burst (peak flux interval and total fluence
interval) are available, generally from four different spectral models. For a
given event, fit results may not be available for all models due to poor
statistics and/or lack of fit convergence.

\begin{figure}[t]
\begin{center}
\includegraphics*[bb=160 50 518 738,height=\textwidth,angle=90]
{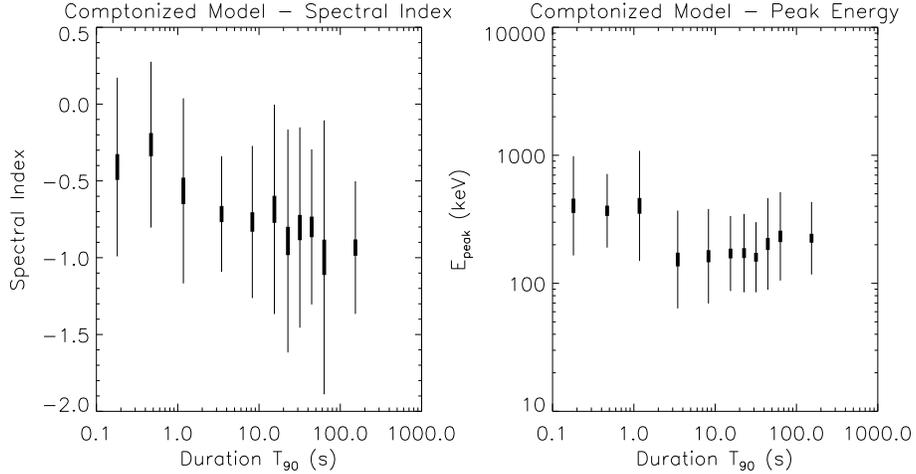}
\end{center}
\caption[]{Spectral parameters as a function of burst duration for fits of a
Comptonized model (power law with exponential high-energy cut-off) using {\it
CONT} data (see text). Thin vertical bars show the width of a Gaussian fit to
the parameter distribution within a duration bin and thick vertical bars show
the error in the mean of each distribution.}
\label{eps1}
\end{figure}

We extracted spectral parameters for all GRBs in the {\it CONT} database for
three of the models (cut-off power law, broken power laws, and the Band GRB
function). For short events, there is little difference between the peak flux
and fluence intervals because of the 2~s {\it CONT} time resolution, whereas
long GRBs typically have harder spectra at the time of peak flux. We binned the
results for each spectral fit parameter according to the burst duration. Since
the distribution of spectral parameters within each duration bin is broad and
approximately Gaussian,  we computed the centroid and width of the best-fitting
Gaussian for each duration bin. Figure~\ref{eps1} shows an example of the
parameters for the cut-off power law model fit to the peak flux intervals,
plotted as a function of burst duration. The left panel shows the power law
spectral index and the right panel shows the cut-off energy. Within a given
duration interval, the thin vertical bars show the width of a Gaussian fit to
the parameter distribution, and the thick vertical bars show the error in the
mean of the distribution. Although the distributions are broad, there appear to
be differences in the trend of the parameters with GRB duration. The hardening
trend in the power law index is roughly continuous throughout, whereas the
trend in the cut-off energy appears more like a step-function, with a
discontinuity around a duration of 2~s, consistent with the minimum in the
$T_{90}$ duration distribution.

Although no quantitative analysis of the statistical significance of these
results has yet been done, distributions of fit parameters for the other models
show essentially the same trends for both peak-flux and fluence intervals. (The
broken power law and Band GRB function fits provide a third parameter, the high
energy power law index, but the statistics of this parameter are not yet good
enough to define clearly its trend with duration.)

It would appear from the right panel of Figure~\ref{eps1} that the energy
spectra of short and long GRBs have different characteristic break energies
that otherwise depend only weakly on duration. Since the break energy is
affected by the Lorentz factor of the expanding fireball as well as by the
redshift of the emitting source, this places interesting limits on the nature
of the sources. Either the short GRBs have higher bulk Lorentz factors or they
are located closer to us than long GRBs, or both. Discovering optical
counterparts for short GRBs and measuring their redshifts would clearly help
resolve the nature of these sources.

%


\begin{thebibliography}{8.}
\addcontentsline{toc}{section}{References}

\bibitem{belli95} B.M. Belli: Ap\&SS \textbf{231}, 43 (1995)

\bibitem{cohen97} E. Cohen, J.I. Katz et al.: ApJ \textbf{488}, 330 (1997)

\bibitem{crider97} A. Crider, E.P. Liang et al.: ApJ \textbf{479}, L39 (1997)

\bibitem{dezalay92} J.-P. Dezalay, C. Barat et al.:
In: \emph{Gamma-Ray Bursts: Huntsville, AL 1991}, ed. by W.S. Paciesas, G.J.
Fishman (AIP Conf.\ Proc.\ 265, New York 1992) pp. 304--309

\bibitem{fenimore96} E.E. Fenimore, C. Madras et al.: ApJ \textbf{473}, 998
(1996)

\bibitem{horvath98} I. Horv\'{a}th: ApJ \textbf{508}, 757 (1998)

\bibitem{kouvel93} C. Kouveliotou, C.A. Meegan et al.: ApJ \textbf{413}, L101
(1993)

\bibitem{mallozzi98} R.S. Mallozzi, G.N. Pendleton et al.: 
In: \emph{Gamma-Ray Bursts: 4th Huntsville Symposium}, ed. by C.A. Meegan, R.D.
Preece, T.M. Koshut (AIP Conf.\ Proc.\ 428, New York 1998) pp. 273--277

\bibitem{mukherjee98} S. Mukherjee, E.D. Feigelson et al.: ApJ \textbf{508}, 314
(1998)

\bibitem{nor2} J.P. Norris, G.F. Marani et al.: ApJ \textbf{534}, 248 (2000)

\bibitem{nor3} J.P. Norris, J.D. Scargle et al.: {\it these proceedings}

\bibitem{preece98a} R.D. Preece, G.N. Pendleton et al.: ApJ \textbf{496}, 849
(1998)

\bibitem{preece98b} R.D. Preece, M.S. Briggs et al.: ApJ \textbf{506}, L23
(1998)

\bibitem{preece00} R.D. Preece, M.S. Briggs et al.: ApJS \textbf{126}, 19 (2000)

\bibitem{preece01} R.D. Preece, M.S. Briggs et al.: {\it in preparation} (2001)

\bibitem{saripiran97} R. Sari, T. Piran: ApJ \textbf{485}, 270 (1997)

\bibitem{tavani96a} M. Tavani: Phys.\ Rev.\ Lett.\ \textbf{76}, 3478 (1996)

\bibitem{tavani96b} M. Tavani: ApJ \textbf{466}, 768 (1996)

\bibitem{tavani98} M. Tavani: ApJ \textbf{497}, L21 (1998)

\end{thebibliography}
\end{document}